# Femtosecond control of terahertz spin-charge conversion in ferromagnetic heterostructures

Xiaojun Wu[1,6*†], Tianxiao Nie[3,4 †], Bo Wang[2,5†], Deyin Kong[1,6], Yang Gao[1,6], Meng Xiao[1], Chandan Pandey[3], Lianggong Wen[3,4], Weisheng Zhao[3,4*], Cunjun Ruan[1,6], Jungang Miao[1,6], Li Wang[2*], Yutong Li[2,5], and Kang L. Wang[7]

**Affiliations:**

[1]School of Electronic and Information Engineering, Beihang University, Beijing, 100191, China

[2]Beijing National Laboratory for Condensed Matter Physics, Institute of Physics, Chinese Academy of Sciences, Beijing 100190, China

[3]Fert Beijing Institute, BDBC, and School of Microelectronics, Beihang University, Beijing 100191, China

[4]Beihang-Goertek Joint Microelectronics Institute, Qingdao Research Institute, Beihang University, Qingdao, 266000, China

[5]School of Physical Sciences, University of Chinese Academy of Sciences, Beijing 100049, China

[6]Beijing Key Laboratory for Microwave Sensing and Security Applications, Beihang University, 100191, China

[7]Device Research Laboratory, Department of Electrical Engineering, and Department of Physics and Astronomy, University of California, Los Angeles, 90095, USA

*To whom correspondence should be addressed. E-mail: xiaojunwu@buaa.edu.cn; weisheng.zhao@buaa.edu.cn; wangli@iphy.ac.cn

†Equal contribution.

**Abstract**: Employing electron spin instead of charge to develop spintronic devices holds the merits of low-power consumption in information technologies. Meanwhile, the demand for increasing speed in spintronics beyond current CMOS technology has further triggered intensive researches for ultrafast control of spins even up to unprecedent terahertz regime. The femtosecond laser has been emerging as a potential technique to generate an ultrafast spin-current burst for magnetization manipulation. However, there is a great challenge to establish all-optical control and monitor of the femtosecond transient spin current. Deep insights into the physics and mechanism are extremely essential for the technique. Here, we demonstrate coherently nonthermal excitation of femtosecond spin-charge current conversion parallel to the magnetization in W/CoFeB/Pt heterostructures driven by linearly polarized femtosecond laser pulses. Through systematical investigation we observe the terahertz emission polarization depends on both the magnetization direction and structural asymmetry. We attribute this phenomenon of the terahertz generation parallel to the magnetization induced by linearly polarized femtosecond laser pulses probably to inverse spin-orbit torque effect. Our work not only is beneficial to the deep understanding of spin-charge conversion and spin transportation, but also helps develop novel on-chip terahertz opto-spintronic devices.

## Main

Femtosecond generation and control of spin currents in solids have received significant attention recently due to its promising capability to enhance the operation speed of spin device up to terahertz (THz) frequency ranges[1–6]. Previously, the interactions between femtosecond laser pulses and solids like semiconductors and magnetics to realize ultrafast control of spins were extensively investigated using transient magneto-optical Kerr microscopy and many other methods[7]. According to the Maxwell's equation, femtosecond transient currents will radiate electromagnetic waves in the THz frequency range[8]. Hence, characterization of the emitted THz waves can be used to deduce the ultrafast femtosecond spin dynamics, which will have a significant impact for the realization of the ultrafast manufacture of electron spins[9]. In this case, THz emission spectroscopy works as a contactless, highly sensitive, time-resolved, coherent "Ampere" meter. Besides, the ferromagnetic (FM)/heavy metal (HM) heterostructures have been developed as highly-efficient, cost-effective, ultrabroadband THz sources, enabling applications in both weak-field spectroscopy and strong-field nonlinear investigations[10–19].

The current reported THz radiation from FM/HM heterostructure nanofilms illuminated by femtosecond laser pulses has been classified into thermal and nonthermal processes, corresponding to helicity-independent and helicity-dependent THz radiation, respectively[8,9]. Both are easily identified by the generated THz electric fields, being either perpendicular or parallel to the magnetization direction. The former is attributed to the presence of laser heating induced spin currents, which flow along the pump laser propagation direction[8,10]. Once the spin current flows from an FM into an HM, it converts to an in-plane transverse charge current perpendicular to the magnetization via inverse spin Hall Effect (ISHE), which in turn results in radiating THz transients. The converted charge current may be expressed by

$$\vec{j}_c^\perp = \theta \vec{j}_s \times \vec{M} / |\vec{M}| \tag{1}$$

where $\vec{j}_c^\perp$ is the converted charge current perpendicular to the magnetization; $\theta$ is the magnitude of spin Hall angle with positive (i.e. Pt, Ir, Mn) or negative (i.e. W, Cr, Ta) values determined by the type of the HM[11]; $\vec{j}_s$ is the injected spin current density determined by the pump laser fluence; $\vec{M}$ is the magnetization. Consequently, the radiation process depends on the pump femtosecond laser pulse intensity, the magnetization and the spin Hall angles. The THz polarization is controlled by the magnetization, but insensitive to the helicity of the pump laser pulses.

However, recent advancements in both bilayer ferromagnetic heterostructures of Co/Pt[9] and Ag/Bi Rashba interfaces[12] have shown an in-plane photocurrent component parallel to the magnetization, which also radiates THz waves. Previously, this phenomenon was interpreted with the help of inverse spin-orbit torque (I-SOT) effect[20], which transfers the tilting of magnetization into the photocurrent transient[9]. Phenomenologically, this magnetization tilting is due to an effective magnetic field induced by either inverse Faraday effect [12] or optically-induced spin transfer torque from a femtosecond circularly polarized light. The induced photocurrent transient by I-SOT from a circularly polarized light can be described as

$$\vec{j}_c^{\parallel} = \chi \hat{n} \times \left[ \vec{M} \times \hat{\sigma} \right] I \qquad (2)$$

where $\vec{j}_c^{\parallel}$ is the induced charge current parallel to the magnetization direction; $\chi$ is a scalar; $\hat{n}$ is the polar unit vector normal to the interface; $\hat{\sigma}$ is the helicity vector of the circularly polarized light; $I$ is the pumping intensity. From this formula, it can be seen that the charge current parallel to the magnetization direction has emission symmetry properties determined by the direction of structural asymmetry (illumination faces of either the FM or the HM), the magnetization direction, as well as the helicity of the pumping laser pulses.

In this work, we demonstrate, to the best of our knowledge, the first time of discovering the femtosecond spin converted photocurrent parallel to the magnetization in CoFeB/HM heterostructures, which are driven by linearly polarized femtosecond laser pulses via I-SOT effect. Furthermore, comparing the results from thickness-dependent W/CoFeB or Pt/CoFeB bilayers, we find that the intentionally designed trilayer structures with large spin-Hall angles but opposite signs of W and Pt can enhance the THz emission parallel to the magnetization. We believe that our findings not only help deepen the physical understanding of femtosecond spin dynamics[21–23], spin transportations[24,25] and ultrafast demagnetization[26–28], but also open the doors for developing novel on-chip THz functional devices and platforms for the next generation of THz opto-spintronics[29–31].

Fifteen samples tested in our experiments are FM/HM and HM/FM/HM heterostructures grown on 0.5 mm thick fused-quartz substrates, in which the FM is CoFeB while the HMs are W and Pt layers. The sample growth was conducted in a high-vacuum AJA sputtering system with a base pressure of $10^{-9}$ Torr. Before the deposition, the substrate was cleaned by Ar plasmonic etching for a dust-free surface. The deposition conditions (i.e. Ar pressure and applied power) were carefully optimized to promise the best quality and reproducibility. The growth rate for W, CoFeB and Pt was 0.21 Å/s, 0.06 Å/s and 0.77 Å/s, respectively. During the deposition, a sample rotation was performed to ensure a good uniformity. Three W/CoFeB/Pt trilayer heterostructures were grown with different thickness. The HM capping layers of W and Pt are both 1.8 nm thick, while the FM layers are 1.8, 2.0, and 2.2 nm, respectively. Ten bilayer samples are W/CoFeB and Pt/CoFeB with fixed FM layer thickness of 2.0 nm, but the HM layers for W and Pt include the thickness of 1.8, 2.0, 2.2, 3, and 4 nm. There is another W/CoFeB bilayer with 5 nm thick W layer on CoFeB. The magnetization of the samples is in-plane, and the magnitude of ~50 mT is set by a static external magnetic field.

The THz emission spectroscopy used in our experiments is driven by a commercial Ti:sapphire laser oscillator delivering pumping pulses with central wavelength of 800 nm, pulse duration of 70 fs, repetition rate of 80 MHz. As shown in FIG. , the laser pulses are divided into two beams. One beam with 90% energy is used either for pumping a ZnTe crystal as the THz emitter or driving the ferromagnetic metal materials for THz emission. The other beam with lower energy is used to probe the generated THz pulses. The focal length used for THz emission is 150 mm. The polarizer $P_2$ is mounted onto an electrically controlled rotation stage, which guarantees accurate measurement of the two transmitted THz electric field components for the ±45° with respect to the vertical direction in the laboratory coordinate. $P_1$ is used to obtain pure incident THz wave polarized vertically for the ZnTe emission spectroscopy, while $P_3$ is applied to insure the same response function for the two THz electric components. After these polarizers, the THz pulses are detected with electro-optic sampling method, which includes a 1 mm thick ZnTe crystal, a λ/4 wave plate, a Wollaston prism and a pair of photodiodes. The THz electric field induces the variation of the refractive index of the detection crystal via linear electro-optic effect (Pockels effect), which is recorded via the difference detection in the photodiodes by a probing beam.

FIG.a shows directly the observed THz electric fields parallel to the magnetization in FM/HM heterostructures. Firstly, we perform the THz emission spectroscopy measurements on the W/CoFeB, Pt/CoFeB bilayers, and W/CoFeB/Pt trilayers driven with linearly polarized laser pulses. For the bilayer samples, we detect a single-cycle $E_x$ signal perpendicular to the magnetization direction of the Pt(2.2)/CoFeB(2.0) (The number inside the brackets indicates the corresponding film thickness in nanometer) bilayer, while no detectable $E_y$ that parallels to the magnetization is obtained, as shown in FIG.b. The $E_x$ signal observed in W(2.2)/CoFeB(2.0) is ~7 times larger than that in the Pt(2.2)/CoFeB(2.0) bilayer, and we also observe a $E_y$ signal in W(2.2)/CoFeB(2.0), as illustrated in FIG.c. The larger $E_x$ from the W/CoFeB bilayer is probably due to larger spin Hall angle of W[32], and the different fabrication conditions[13]. The undetected $E_y$ in the Pt/CoFeB bilayer is beyond the machine's suitability. Furthermore, the thickness-dependent $E_x$ in HM is also investigated, and the results show that the optimal thickness for W is equal to 2.2 nm. Such behavior agrees well with the already reported results in Ref. [11], indicating the similar mechanism for $E_x$ emission. We did not detect any THz emission from the CoFeB single layer on glass substrate due to the negligible ISHE.

Interestingly, from a trilayer W(1.8)/CoFeB(2.0)/Pt(1.8), we observe $E_x$ signal more than 4 times higher than those of the W(2.2)/CoFeB(2.0) bilayer, as shown in FIG.d[11]. Similar results are also observed in the other two trilayers of W(1.8)/CoFeB(1.8)/Pt(1.8) and W(1.8)/CoFeB(2.2)/Pt(1.8). The effect is possibly for engineering of structures to significantly boost the THz emission. Our data can be attributed to the in-phase enhancement in the opposite spin-Hall angles of W and Pt layers. The best optimal efficiency is one fourth of that of a 1 mm thick ZnTe emitter under the same pumping parameters. With 10 fs laser pumping[11], the THz efficiency in our trilayer structure would be much closer to that from a ZnTe emitter. The reason is that the longer pump pulse duration of 70 fs in our case leads to a longer effective interaction length and to a higher efficiency in ZnTe due to optical rectification, while with 10 fs laser pulse pumping, the optical rectification is suppressed by the short effective interaction length. We have already shown very high THz efficiency and the feasibility of using FM/NM heterostructures as THz sources driven by high repetition rate femtosecond laser pulses with moderate pulse durations.

Besides the highly efficient THz generation whose polarization is perpendicular to the magnetization in W/CoFeB/Pt trilayers, we detect clearly resolved $E_y$ components, as illustrated in FIG.d. The maximum temporal waveform amplitude ratio of $E_y/E_x$ is 0.05, which is comparable to most of the reported values from Co/Pt heterostructures in Ref. [9]. The peak times of the $E_x$ and $E_y$ are not rigorously synchronized, which causes the output THz waves to be elliptically polarized. To the best of our knowledge, this is the first observation of the parallel magnetization induced THz radiation in CoFeB related heterostructures by femtosecond pulses[12]. This phenomenon has also been repeated in the other two trilayers of W(1.8)/CoFeB(1.8)/Pt(1.8) and W(1.8)/CoFeB(2.2)/Pt(1.8).

In order to investigate the origin of the $E_y$ component that is paralleled to the magnetization, we systematically study the emitted THz polarity dependence on the illumination directionality and the magnetization direction in the W(1.8)/CoFeB(2.0)/Pt(1.8) trilayer. FIG. a and d show the schematic diagrams used in symmetric geometry studying the magnetization reversal induced THz polarity. FIG. b shows the dependence of $E_y$, $\vec{M}$ and $\hat{n}$ for the W/CoFeB/Pt and Pt/CoFeB/W samples, pumped by linearly polarized laser pulses. The peak-to-peak amplitudes for both incident directions are almost the same. The observed directional effects rule out the femtosecond demagnetization as the primary mechanism for the $E_y$ generation, because femtosecond demagnetization should have dependence of the directionality[17,33]. FIG. e exhibits the THz emission polarity reversal behavior for $E_y$ when we reverse the external applied magnetic field direction.

Similar results have also been obtained in the W/CoFeB/Pt trilayers, as shown in FIG. c and f. The $E_x$ component of the THz emission demonstrates a sign change as the polar vector $\hat{n}$ and magnetization $\vec{M}$ are reversed. Together with the spin Hall angles dependence, these phenomena give a clue that the origin of the $E_x$ emission may arise from the electric current emerged from the ISHE generated from spin current induced by a femtosecond laser pulse.

We include that heat-assisted ISHE cannot be used to explain the generation of THz $E_y$ signals parallel to the magnetization because the emitted THz signals in such effect should always be linearly polarized with its polarization perpendicular to the magnetization according to equation (1)[8,12,15]. Therefore, we argue the possible mechanism of our THz generation parallel to the magnetization driven by linearly polarized femtosecond laser pulses may be due to I-SOT effect[20,34]. The I-SOT could induce an effective magnetic field to tilt the magnetization which in turn gives a photocurrent induced THz emission independent on laser helicity.

To gain deep insight into the physics of the $E_y$ generation, we perform the polarization dependent THz emission on the W(1.8)/CoFeB(1.8)/Pt(1.8) trilayer. FIG. give the pump laser polarization dependent $E_x$ and $E_y$ polarity for different magnetization direction. For different polarization of the pump laser, both of the $E_x$ and $E_y$ components preserve the phase property, and there are no remarkable differences. However, all the $E_y$ components appear ~0.3 ps earlier of than the $E_x$ components. It implies that the effective refractive index along with the external magnetic fields is smaller than that for the perpendicular case.

Although the detected values of $E_y$ is small, we can still coherently combine it with $E_x$ to form elliptical THz beams. FIG. a shows the typical single-cycle $E_x$ and $E_y$ THz components in the W(1.8)/CoFeB(2.0)/Pt(1.8) pumped from the W side ($n^+$). The $E_y$ signal is magnified by 5 times for clarity. The corresponding Fourier transform spectra as well as the phases of $E_x$ and $E_y$ obtained in our measurements are illustrated in FIG. b. The detected THz frequency range is in 0.2-2.8 THz, which is limited by the pump laser pulses (100 fs in our case). FIG. c shows the elliptical THz beam in the time domain. The ellipticity dispersion which is shown in FIG. d slightly increases along with the frequency for the W/CoFeB/Pt ($n^+$). The phase difference and the ratio of the amplitudes all contribute to the ellipticity and there may be a mechanism that not only related to the phase but also relevant to the amplitude ratio. When the $E_x$ and $E_y$ components have time delay, the phase difference varies from 0 rad (or $\pi$ rad when arbitrary component is reversed), and it is directly proportional to the frequencies. We use function $y = ax + b$ to fit the original phase difference from ~0.5 THz to ~2.3 THz for the high signal to noise ratio frequency range. The slope is -0.21 rad/THz and the intercept is ~2.15 rad which means the phase difference is not simply influenced by the time difference. The paradoxes between time delay due to refractive index difference for $E_x$ and $E_y$ and the extracted phase difference implies that the microscopic origin of the $E_y$ is ill-defined and needs further investigations. The deviation may be caused by the complex dielectric constant of the metal. The complex refractive index can be calculated by the equation

$$\dot{n} = \sqrt{\dot{\varepsilon}_{metal}} = \sqrt{\varepsilon_\infty + i\frac{\dot{\sigma}_{metal}}{\omega \varepsilon_0}} \tag{3}$$

where $\dot{n}$ is the complex refractive index, $\dot{\varepsilon}_{metal}$ the complex dielectric constant of the metal, $\varepsilon_\infty$ the permittivity at infinite frequency, $i$ the imaginary unit, $\dot{\sigma}_{metal}$ the complex conductivity of the metal, $\omega$ the angular frequency, and $\varepsilon_0$ the dielectric constant in vacuum. When the frequency decrease, the real refractive index will increase, and this may make the phase difference not directly proportional to the frequency.

In order to estimate the induced photocurrent density in W/CoFeB/Pt, we use the Maxwell's equations and some approximations as shown in Supplementary Information. For such estimation, we assume the femtosecond laser pulses are illuminated from the W layer ($n^+$) side, and the currents are originated from the inversion breaking geometry which only occurs at the interfaces with a thickness less than 0.2 nm. Meanwhile, we also assume the two photocurrent components with their density along the x and y axis are in the form of the Gaussian function with a variable width and amplitude, since the THz time-domain spectrometer is a broadband system. The width of this Gaussian function gives the detected spectral bandwidth and represents the timescale in which the dynamics occur, while the amplitude of this function determines the observed electric field amplitude and represents the maximum current density amplitude. For the measured THz electric field components, the single pulse energy is in the order of fJ-level from the W/CoFeB/Pt trilayers, which is comparable to that from ZnTe. The estimated $E_x$ is ~1 V/cm, while $E_y$ is ~0.1 V/cm for both maximum THz emission signals. We take the conductivity of this heterostructure to be on the order of $10^5 \Omega^{-1} m^{-1}$, and obtain an estimate of the y-component current pulse in the W/CoFeB/Pt heterostructure[11]. The maximum current density amplitudes for the interfaces of the W/CoFeB and CoFeB/Pt are thus estimated to be on the order of $10^7 \, Am^{-2}$ and $10^5 \, Am^{-2}$, respectively.

FIG. shows the dependence of THz yields on pump fluence dependent. The THz amplitude is defined as the peak-to-peak signals illustrated in FIG.d. The $E_y$ signals from the W/CoFeB/Pt ($n^-$) is slightly larger than those obtained from the Pt/CoFeB/W ($n^+$) when the pump fluence is higher than 1.0 μJ/cm². (see FIG. a). Later, the amplitudes start to saturate when further increasing the pump fluence. Although we use a Ti:sapphire oscillator (nJ single pulse energy), rather than an laser amplifier (mJ) for the pump laser pulses, the average powers are also at the microwatt level. Therefore, femtosecond laser heating can induce conductivity variation and the decrease of magnetization can also happen in our cases contributing to the saturation behavior[9]. Other possible saturation mechanism may originate from spin current saturation arising from the spin accumulation in the HM layers. For the $E_x$ component, the THz signals obtained from the W/CoFeB/Pt ($n^+$) are always weaker than that from the Pt/CoFeB/W ($n^-$), and we also observe slightly saturation along with increasing pump fluence (see FIG. b). For the $n^+$ case, the THz wave is first generated in the heterostructure and then propagates through the glass substrate. The refractive index for the glass substrate is ~1.963 (1 THz)[35]. The Fresnel reflection loss is ~10%. The absorption coefficient is ~2.6 cm$^{-1}$ (1 THz). Due to Fresnel reflection loss and absorption of the substrate, the out-coupled THz signals are ~15% smaller than those in the case of $n^-$ for the pump fluence of 2.5 μJ/cm$^2$, of which there is no such losses because the THz waves are directly out-coupled into free-space from the heterostructure.

We have demonstrated the generation of THz radiation induced by nonthermal femtosecond laser induced spin converted photocurrent. The THz electric fields as shown to be parallel to the magnetization in the W/CoFeB/Pt trilayer heterostructures driven by linearly polarized femtosecond pulses. The origin of the photocurrent is shown to arise from the I-SOT. Through systematical investigations between trilayer and bilayer CoFeB-related heterostructures, the specially designed two HM layers will opposite spin Hall angles are shown to enhance the femtosecond photocurrents due to I-SOT effect parallel to the magnetization. This effect has been not observed before or neglected previously. The observation is not only of significance for extending fundamental research of nanospintronics to THz frequencies, but also opens a door towards understanding all-optical control of magnetization and advancing future ultrafast magnetic recording and information technologies. With the spin-orbit interaction and using special asymmetric geometries, novel integrated functional on-chip THz devices and systems can be developed, accelerating the advancement and applications of ultrafast THz opto-spintronics.

**Methods**

1. **Photocurrent estimation from Maxwell equation**

In this part, we derive mathematical expressions describing how a fast laser pulse induces the emitted THz radiation. We assume a heterostructure in the $xz$ plane as indicated in Fig. M1[9].

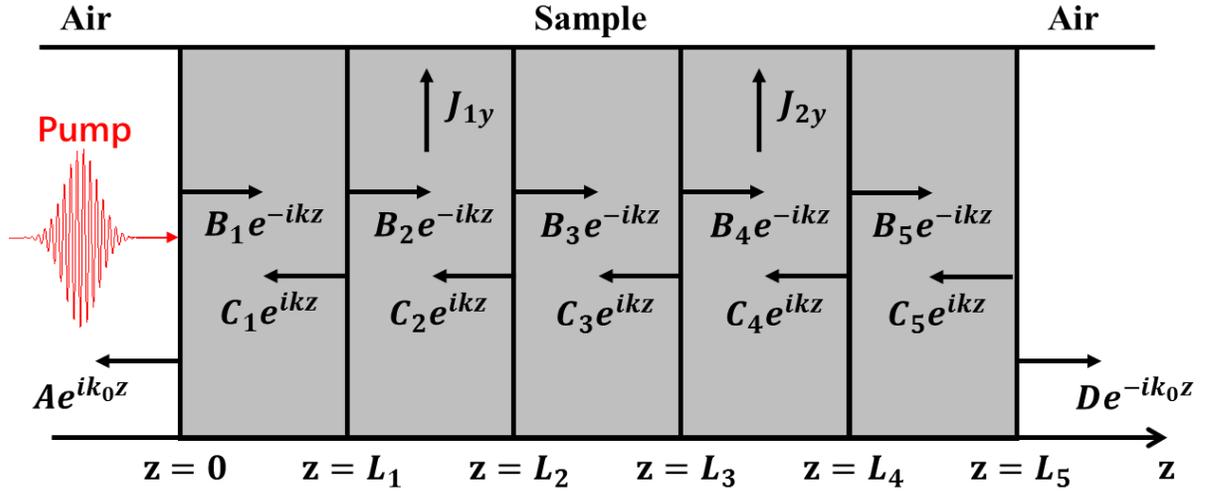

**Fig. M1.** Schematic of the heterostructure in the $xz$ plane with $L_n = \sum_{m=1}^{m=n} l_m$.

As for $E_y$ which is parallel with respect to the magnetization, we use Ampere's and Faraday's laws to relate the electric field to the current of the medium, which in Gaussian units and in frequency domain can be expressed as:

$$\frac{\partial^2 E_y(z)}{\partial z^2} + \frac{\omega^2}{c^2}\varepsilon(z)E_y(z) = -\frac{4\pi i\omega}{c^2}J_y(z) \tag{M1}$$

where $E_y$ is the x-component of the electric field, $\omega$ is the angular frequency, $c$ is the speed of light in vacuum, $\varepsilon(z)$ is the dielectric permittivity, $J_y$ is the y-component of the current density and the ~ symbol is used to indicate the Fourier transform with respect to time.

As shown in Fig. M1, we assume the five metallic layers have approximately the same index of refraction in which the second layer and the fourth layer represent the interfaces exhibiting current dynamics and the two outmost layers are adjacent to vacuum or air. So, we can write the current density as follows:

$$J_{1y}(\omega, z) = J_{1y}(\omega)[u(z - l_1) - u(z - l_1 - l_2)] \tag{M2a}$$

$$J_{2y}(\omega, z) = J_{2y}(\omega)[u(z - l_1 - l_2 - l_3) - u(z - l_1 - l_2 - l_3 - l_4)] \tag{M2b}$$

where $u$ is the Heaviside step function, $l_n$ is the thickness of the n-th metallic layer, with $n = 1, 2, 3, 4, 5$. To get the expressions of the y-component of the electric field, we have to calculate the fundamental solution of the homogeneous equation and a partial solution of inhomogeneous equation (M1). The result can be expressed as follows:

$$E_y(z) = \begin{cases} Ae^{ik_0 z}, z < 0 \\ B_1 e^{-ikz} + C_1 e^{ikz}, 0 < z < l_1 \\ B_2 e^{-ikz} + C_2 e^{ikz} + \phi_1, l_1 < z < (l_1 + l_2) \\ B_3 e^{-ikz} + C_3 e^{ikz}, (l_1 + l_2) < z < (l_1 + l_2 + l_3) \\ B_4 e^{-ikz} + C_4 e^{ikz} + \phi_2, (l_1 + l_2 + l_3) < z < (l_1 + l_2 + l_3 + l_4) \\ B_5 e^{-ikz} + C_5 e^{ikz}, (l_1 + l_2 + l_3 + l_4) < z < (l_1 + l_2 + l_3 + l_4 + l_5) \\ De^{-ik_0 z}, z > (l_1 + l_2 + l_3 + l_4 + l_5) \end{cases} \tag{M3}$$

where $\phi_1$ and $\phi_2$ are partial solutions, $k_0$ is the wave vector in air, the sign of the item $\pm ik_0 z$ is decided by the transmission direction, $k$ is the wave vector in the metallic layers and

$l_n$ is the thickness of the n-th metallic layer, with $n = 1, 2, 3, 4, 5$. Taking the partial solutions as constants, it is found from eq. M1 as $\phi_1 = -\dfrac{4\pi i}{\omega \varepsilon_1} J_{1y} \approx -\sigma_1^{-1} J_{1y}$ and $\phi_2 = -\dfrac{4\pi i}{\omega \varepsilon_2} J_{2y} \approx -\sigma_2^{-1} J_{2y}$, with $\varepsilon_n$ the permittivity of the n-th layer where the currents are located and $\sigma_n$ the corresponding conductivity, in accordance with Ohm's law. From Faraday's law we know that the electric field of eq. (M3) at the interfaces should be continuous. Furthermore, there is no free charge inside the metal conductor, so that the derivative of the electric field at the interfaces should also be continuous. This provides the boundary conditions:

$$\begin{cases} A = B_1 + C_1 \\ B_1 e^{-ikl_1} + C_1 e^{ikl_1} = B_2 e^{-ikl_1} + C_2 e^{ikl_1} + \phi_1 \\ B_2 e^{-ik(l_1+l_2)} + C_2 e^{ik(l_1+l_2)} + \phi_1 = B_3 e^{-ik(l_1+l_2)} + C_3 e^{ik(l_1+l_2)} \\ B_3 e^{-ik(l_1+l_2+l_3)} + C_3 e^{ik(l_1+l_2+l_3)} = B_4 e^{-ik(l_1+l_2+l_3)} + C_4 e^{ik(l_1+l_2+l_3)} + \phi_2 \\ B_4 e^{-ik(l_1+l_2+l_3+l_4)} + C_4 e^{ik(l_1+l_2+l_3+l_4)} + \phi_2 = B_5 e^{-ik(l_1+l_2+l_3+l_4)} + C_5 e^{ik(l_1+l_2+l_3+l_4)} \\ B_5 e^{-ik(l_1+l_2+l_3+l_4+l_5)} + C_5 e^{ik(l_1+l_2+l_3+l_4+l_5)} = D e^{-ik_0(l_1+l_2+l_3+l_4+l_5)} \\ k_0 A = -k B_1 + k C_1 \\ -B_1 e^{-ikl_1} + C_1 e^{ikl_1} = -B_2 e^{-ikl_1} + C_2 e^{ikl_1} \\ -B_2 e^{-ik(l_1+l_2)} + C_2 e^{ik(l_1+l_2)} = -B_3 e^{-ik(l_1+l_2)} + C_3 e^{ik(l_1+l_2)} \\ -B_3 e^{-ik(l_1+l_2+l_3)} + C_3 e^{ik(l_1+l_2+l_3)} = -B_4 e^{-ik(l_1+l_2+l_3)} + C_4 e^{ik(l_1+l_2+l_3)} \\ -B_4 e^{-ik(l_1+l_2+l_3+l_4)} + C_4 e^{ik(l_1+l_2+l_3+l_4)} = -B_5 e^{-ik(l_1+l_2+l_3+l_4)} + C_5 e^{ik(l_1+l_2+l_3+l_4)} \\ -k B_5 e^{-ik(l_1+l_2+l_3+l_4+l_5)} + k C_5 e^{ik(l_1+l_2+l_3+l_4+l_5)} = -k_0 D e^{-ik_0(l_1+l_2+l_3+l_4+l_5)} \end{cases} \quad (M4)$$

Solving these equations for D, which is the y-component of the emitted field amplitude, provides us with

$$D = \\ -\sigma_1^{-1} k J_{1y} \dfrac{(e^{-ik(l_1+l_2)} - e^{-ikl_1})(k - k_0) - (e^{ik(l_1+l_2)} - e^{ikl_1})(k + k_0)}{e^{-i(k+k_0)(l_1+l_2+l_3+l_4+l_5)}(k - k_0)^2 - e^{i(k-k_0)(l_1+l_2+l_3+l_4+l_5)}(k + k_0)^2} \\ -\sigma_2^{-1} k J_{2y} \dfrac{(e^{-ik(l_1+l_2+l_3+l_4)} - e^{-ik(l_1+l_2+l_3)})(k - k_0) - (e^{ik(l_1+l_2+l_3+l_4)} - e^{ik(l_1+l_2+l_3)})(k + k_0)}{e^{-i(k+k_0)(l_1+l_2+l_3+l_4+l_5)}(k - k_0)^2 - e^{i(k-k_0)(l_1+l_2+l_3+l_4+l_5)}(k + k_0)^2} \quad (M5)$$

which is the expression we used for our calculations to estimate the current density. In order to convert the emitted field into the detected values, we have to consider the response of the spectrometer. We use the method described in Ref. [9] and Ref. [36] to estimate our detected electric field in y direction, which corresponds to D.

To estimate the strength of the photocurrent we assume that the current arising from the inversion symmetry breaking occurs only in the interface layers with a thickness of the order of 0.2 nm, corresponding to $l_2 = l_4 = 0.2$ nm. And we took $l_1 = 1.8 nm$, $l_3 = l_5 = 1.6$ nm, which is the thickness of each metal material. Considering that our spectrometer is a broadband system, we assume the two induced photocurrent density are in the form of Gaussian function with a variable width and amplitude. The width of this Gaussian function determines the spectral bandwidth observed and represents the timescale in which the dynamics occur, while the amplitude of this function determines the observed electric field amplitude and represents the maximum current density amplitude. We take the order of $10^5 \Omega^{-1} m^{-1}$ for the conductivity of this heterostructure. Using eq. M5, we arrive at an estimate for the y-component current pulse in the W/CoFeB/Pt heterostructure. The maximum current density amplitude of $J_{1y}$ and $J_{2y}$ are in the order of $10^7 Am^{-2}$ and $10^5 Am^{-2}$, respectively.

## 2. Data analysis of elliptical THz generation

Fig. M2 presents the definition of the coordinate system. $E_1$ and $E_2$ indicate detected THz electric fields and their angles to x-axis are $45°$ and $-45°$, respectively. In this laboratory coordinate system, the electric field has relationship as:

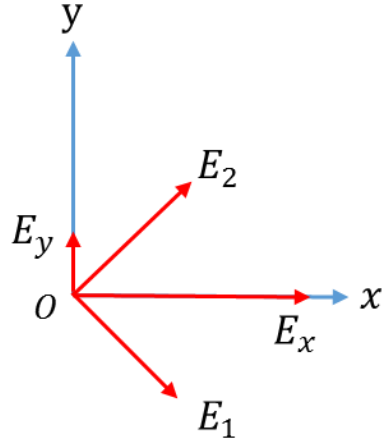

**Fig. M2.** The definition of the coordinate system and the schematic of the electric field.

$$E_x \propto E_1 + E_2 \qquad \text{(M6a)}$$

$$E_y \propto E_2 - E_1 \qquad \text{(M6b)}$$

By performing Fourier transform on $E_x$ and $E_y$, the amplitude $a_1$, $a_2$ and phase $\delta_1$, $\delta_2$ of a single frequency component can be obtained. Under this frequency component, the two electric field components in the laboratory coordinate system can be written as:

$$E_x = a_1 \cos(\omega t + \delta_1) \qquad \text{(M7a)}$$

$$E_y = a_2 \cos(\omega t + \delta_2) \qquad \text{(M7b)}$$

With the knowledge of trigonometric function, it is obvious that

$$-a_1 \leq E_x \leq a_1 \qquad \text{(M8a)}$$

$$-a_2 \leq E_y \leq a_2 \qquad \text{(M8b)}$$

The elliptically polarized light with a frequency of $\omega$ is shown in Fig. M3. $\Psi$ indicates the angle between the long axis of the ellipse and the x-axis, $\eta$ indicates the short axis direction of the ellipse, and $\xi$ indicates the long axis direction of the ellipse.

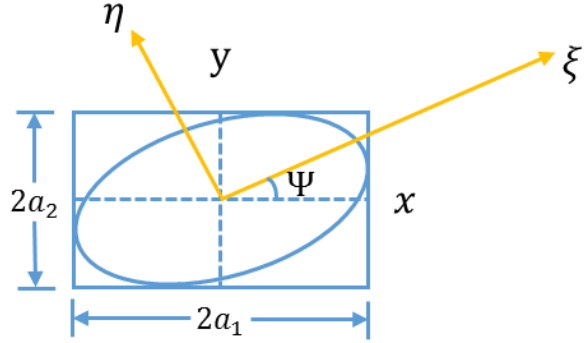

**Fig. M3.** The schematic diagram of the elliptically polarized light.

We assume the amplitude ratio of the electric field parallel to x-axis and y-axis is

$$\frac{a_2}{a_1} = \tan(\alpha) \tag{M9}$$

The phase difference is defined as:

$$\delta_2 - \delta_1 = \delta \tag{M10}$$

It can be derived that the long axis direction of the ellipse satisfies the following equation:

$$\tan(2\Psi) = \tan(2\alpha)\cos(\delta) \tag{M11}$$

The ellipticity is defined as the ratio of the elliptical short semi-axis to the long semi-axis.

$$\mp\frac{b}{a} = \tan(\chi) \tag{M12}$$

where $\mp$ indicates the direction of rotation of the ellipse.

From these equations, we can deduce the expression of $\chi$:

$$\sin(2\chi) = \sin(2\alpha)\sin(\delta) \tag{M13}$$

Through the above analysis, we can obtain $\chi$ and $\Psi$ for each frequency component of the elliptically polarized THz wave using the THz time domain system with polarization measurement.

## 3. Polarization calibration in ZnTe emitter

In order to verify the measurement accuracy of our polarization dependent THz emission spectroscopy, we perform a control experiment in ZnTe emission crystals. We replace ferromagnetic samples with a 1 mm thick ZnTe as the THz emitter. After it, we add a THz polarizer to produce purely vertically polarized THz pulses. The linearly polarized THz pulses first pass through the automatically controlled THz polarizer which records two electric field components at $\pm 45$ degree angles expressed as $E_1$ and $E_2$. Therefore, the THz electric field component of $E_{vertical}$ (in the main text, it is defined as $E_x$ which is perpendicular with respect to the magnetization) and $E_{horizontal}$ (in the main text, it is defined as $E_y$ parallel to the magnetization) can be calculated by $E_1 + E_2$ and $E_2 - E_1$, respectively. Fig. M4 shows the measured vertical and horizontal electric fields in the laboratory coordinate. From this experimental result, we can see that there is absolutely no resolved signal in the horizontal direction, and the largest amplitude of $E_{horizontal}/E_{vertcial} = 0.008$. This value is only one sixth of the result from W/CoFeB/Pt. With this method, we can verify that the observed THz emission component parallel to the magnetization in W/CoFeB/Pt is reasonable.

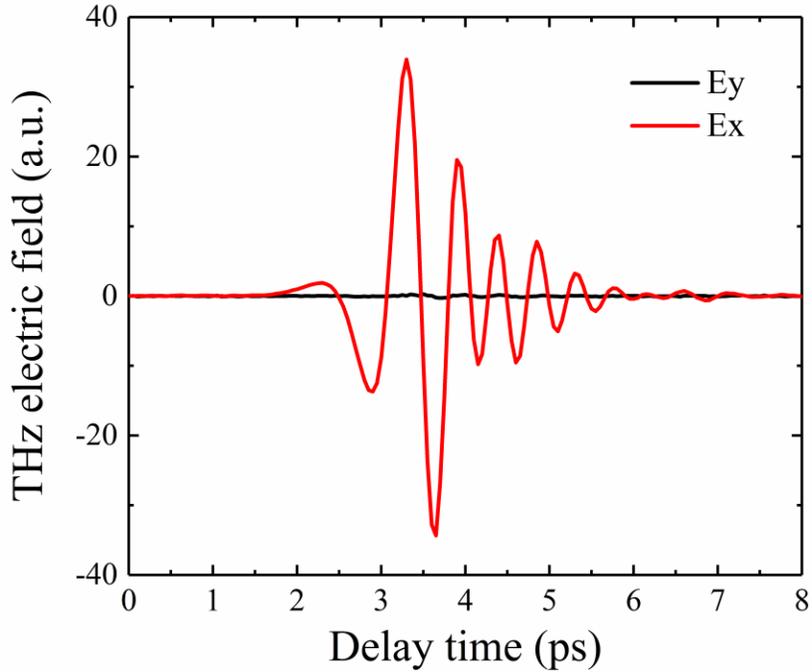

**Fig. M4.** THz emission spectroscopy on ZnTe crystal to verify the measurement accuracy of the polarization dependence.

**Acknowledgments:** This work is supported by the National Nature Science Foundation of China (Grants No. 11827807, 11520101003, 11861121001, 61831001 and 61775233), the Strategic Priority Research Program of the Chinese Academy of Sciences (Grant No. XDB16010200 and XDB07030300), the National Basic Research Program of China (Grant No. 2014CB339800), the International Collaboration Project (Grant No. B16001), and the National Key Technology Program of China (Grant No. 2017ZX01032101). Dr. Xiaojun Wu thanks the "Zhuoyue" Program and "Qingba" Program of Beihang University (Grant No. GZ216S1807, ZG226S1832, KG12052501). Dr. Tianxiao Nie thanks the support from the 1000-Young talent program of China.

**Competing interests**

The authors declare no competing financial interests.

**Author contributions:** X.J.W., and T.X.N. conceived and coordinate the femtosecond control of spin-charge current conversion and terahertz emission project. B.W., D.Y.K., Y.G., and Y.C.G. performed the measurements and analyzed the samples with help from X.J.Wu. M. X. analyzed the data and draw the figures with help from X.J.W. T.X.N. and C.P. designed and fabricated the samples. The theoretical formalisms were derived by M.X., and D.Y.K. with help from X.J.W., and T.X.N. With contributions from C.J.R., W.S.Z., J.G.M., Y.T.L., and K.L.W., X.J.W. wrote the paper. All authors discussed the results and commented on the manuscript.

**Figure legends**

**Fig. 1 Experimental setup of the polarization dependent THz emission spectroscopy.**
$P_1$, $P_2$, $P_3$ are THz polarizers. θ denotes the angle between x-axis and the polarization direction of the excitation laser pulses.

**Fig. 2. Experimental schematic diagram and the layer-dependent THz emission. a,** Experimental schematics and THz emission from W/CoFeB, Pt/CoFeB, and W/CoFeB/Pt heterostructures pumped by linearly polarized nJ femtosecond laser pulses. In the laboratory coordinate, z-axis is the pumping laser propagation direction; y-axis is lying parallel to the magnetization, while x-axis is perpendicular to both the laser propagation and magnetization directions. The laser pulse propagation from W into CoFeB is defined as $n^+$, while from Pt is $n^-$ for the symmetry breaking directionality. The magnetization along with y-axis direction is defined as $M^+$, while the opposite is $M^-$. The radiated THz electric field component parallel and perpendicular to the magnetization is defined as $E_y$ and $E_x$, respectively. **b and c,** Measured THz temporal waveforms for $E_x$ and $E_y$ electric field components in W(2.2)/CoFeB(2.0), and Pt(2.2)/CoFeB(2.0), respectively. **d,** The detected THz signals of $E_y$ and $E_x$ in W(1.8)/CoFeB(2.0)/Pt(1.8) trilayers.

**Fig. 3. Symmetry investigation for THz emission polarity in both $E_y$ and $E_x$ components. a and d,** Schematic diagrams of the THz emission polarity dependent on symmetry breaking

directionality and magnetization direction in W(1.8)/CoFeB(2.0)/Pt(1.8), respectively. $n^+$ means that the laser illuminates on the W surface, while $n^-$ for Pt. $M^+$ means the external applied magnetic field along y axis. **b and c,** symmetry breaking directionality induced THz emission polarity reversal for $E_y$ and $E_x$ when fixing the magnetization direction. **e and f,** THz polarity reversal of $E_y$ and $E_x$ due to reversing the external applied magnetic field direction when the femtosecond laser pulses are illuminated from W facets.

**Fig. 4. The temporal waveforms of the THz signals with different pump laser polarization. a and c,** The $E_y$ components for different pump laser polarization with the laser illumination faces of W and Pt, respectively, while **b and d,** the $E_x$ components that perpendicular to the magnetic field. ⊥: the laser polarization perpendicular to the magnetic field; ∥: the laser polarization parallel to the magnetic field; ↺ and ↻: left-handed and right-handed circularly polarized;

**Fig. 5. Typical combined elliptical THz beam from W/CoFeB/Pt trilayer heterostructures. a,** The temporal waveforms, and **b,** their corresponding Fourier transform spectrum and phase. **c,** Three-dimensional drawing of the combined elliptical THz beam. **d,** Ellipticity, phase difference and ratio of the amplitudes as a function of the generated THz frequency. The straight lines are the linear fitting in the specific frequency range. The polarization of the pump laser is linearly polarized with the direction perpendicular to the magnetic field.

**Fig. 6. Pump fluence dependent THz emission amplitude in W/CoFeB/Pt trilayers.** a. Pump fluence dependence of $E_y$ amplitudes, and b. $E_x$ amplitudes for the pumping incidence from the sides of W layer ($n^+$) and Pt layer ($n^-$), respectively.

**Figures**

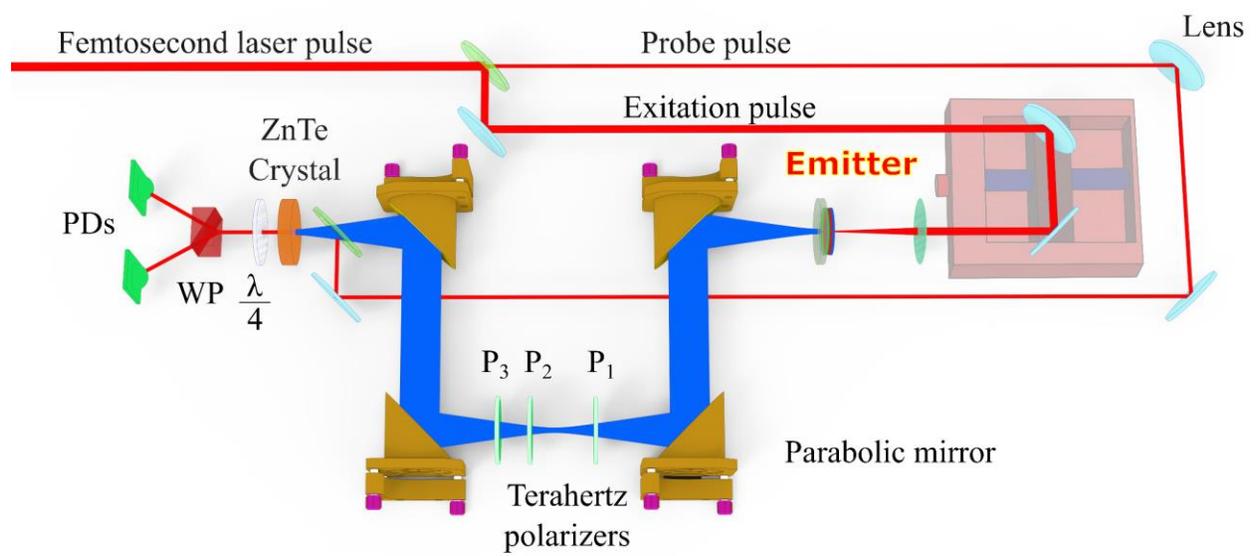

**FIG. 1**

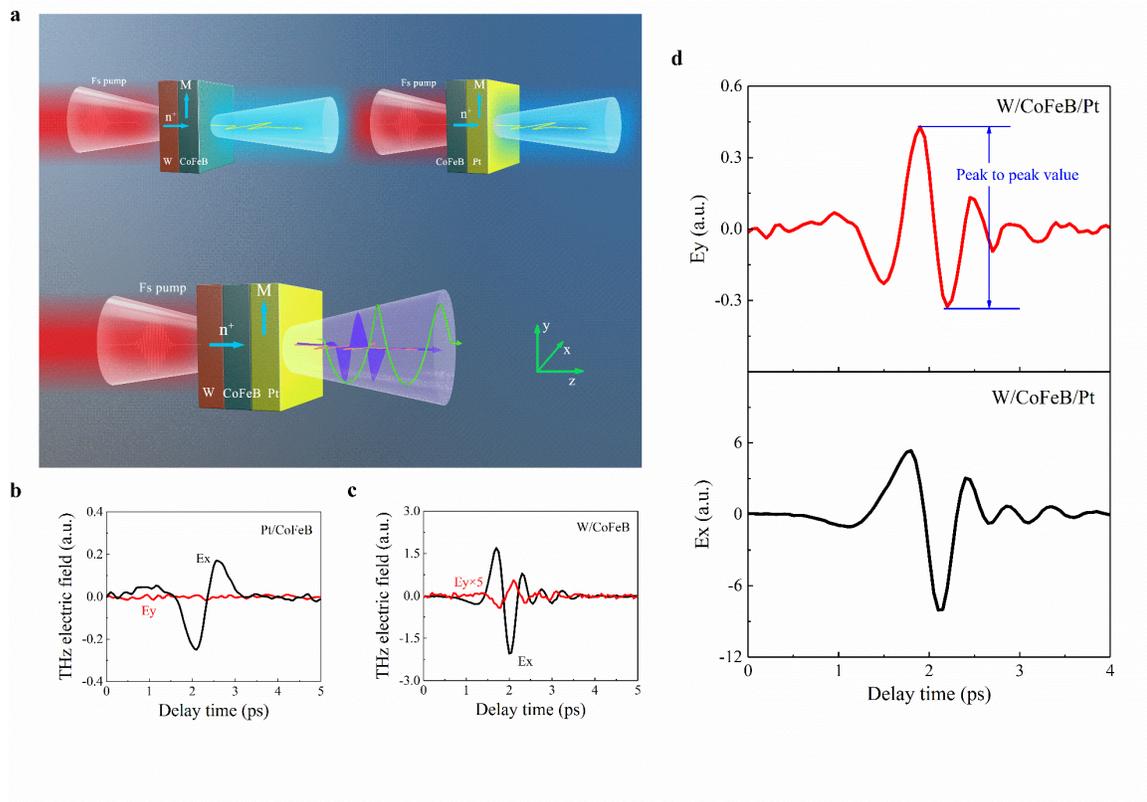

**FIG. 2**

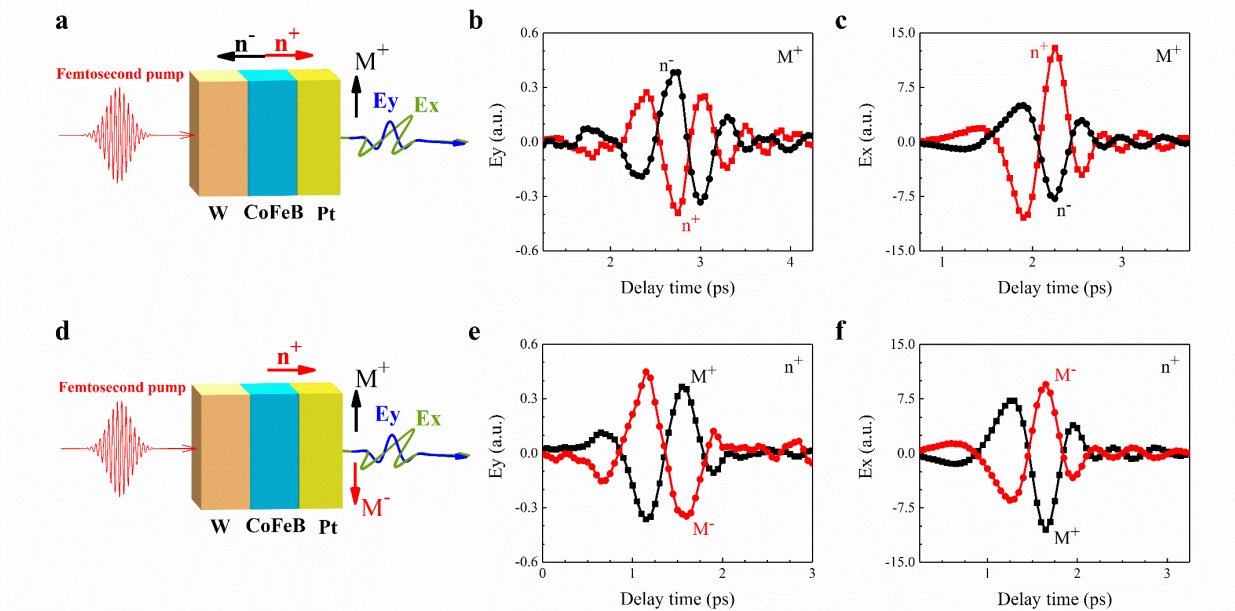

**FIG. 3**

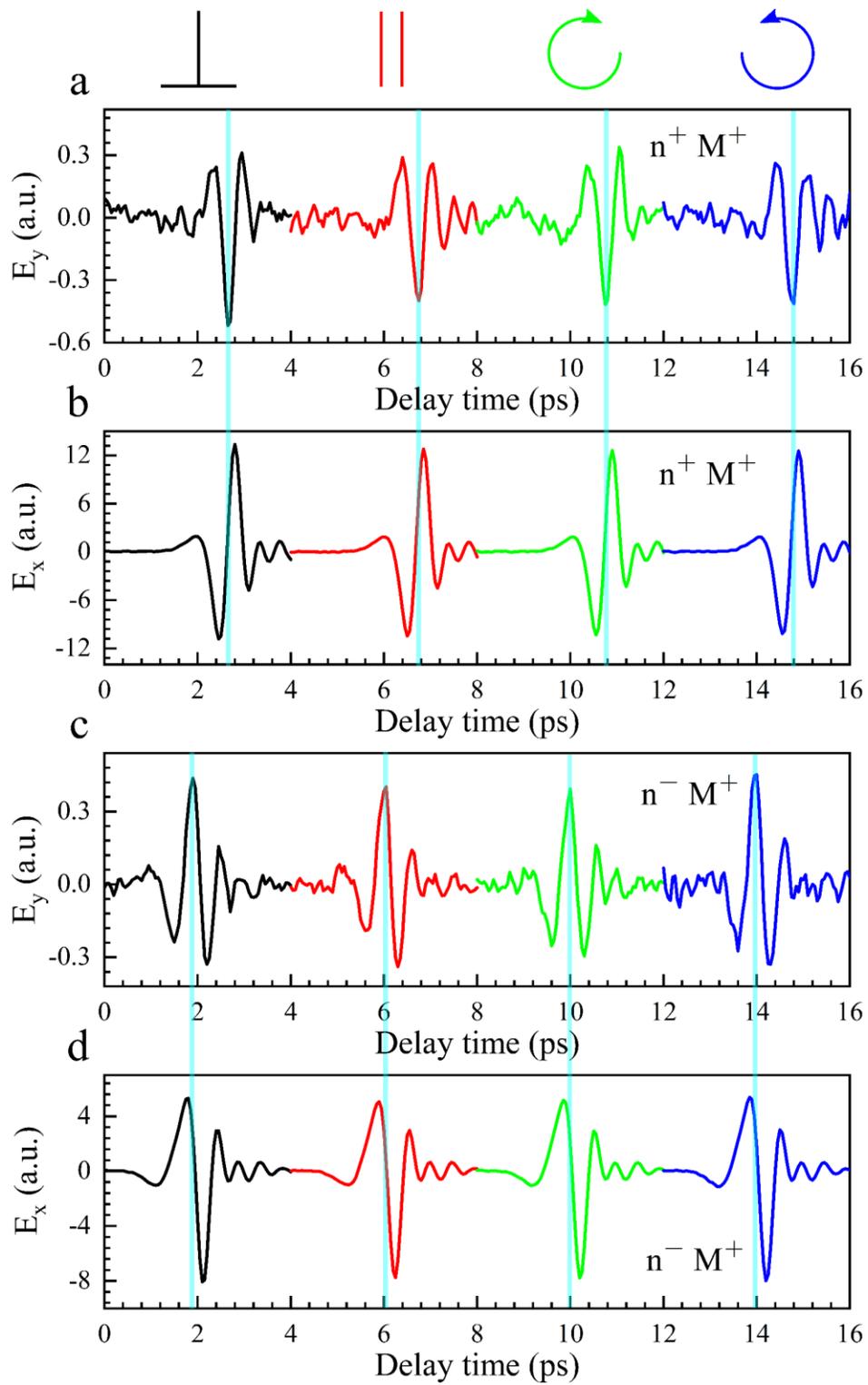

**FIG. 4**

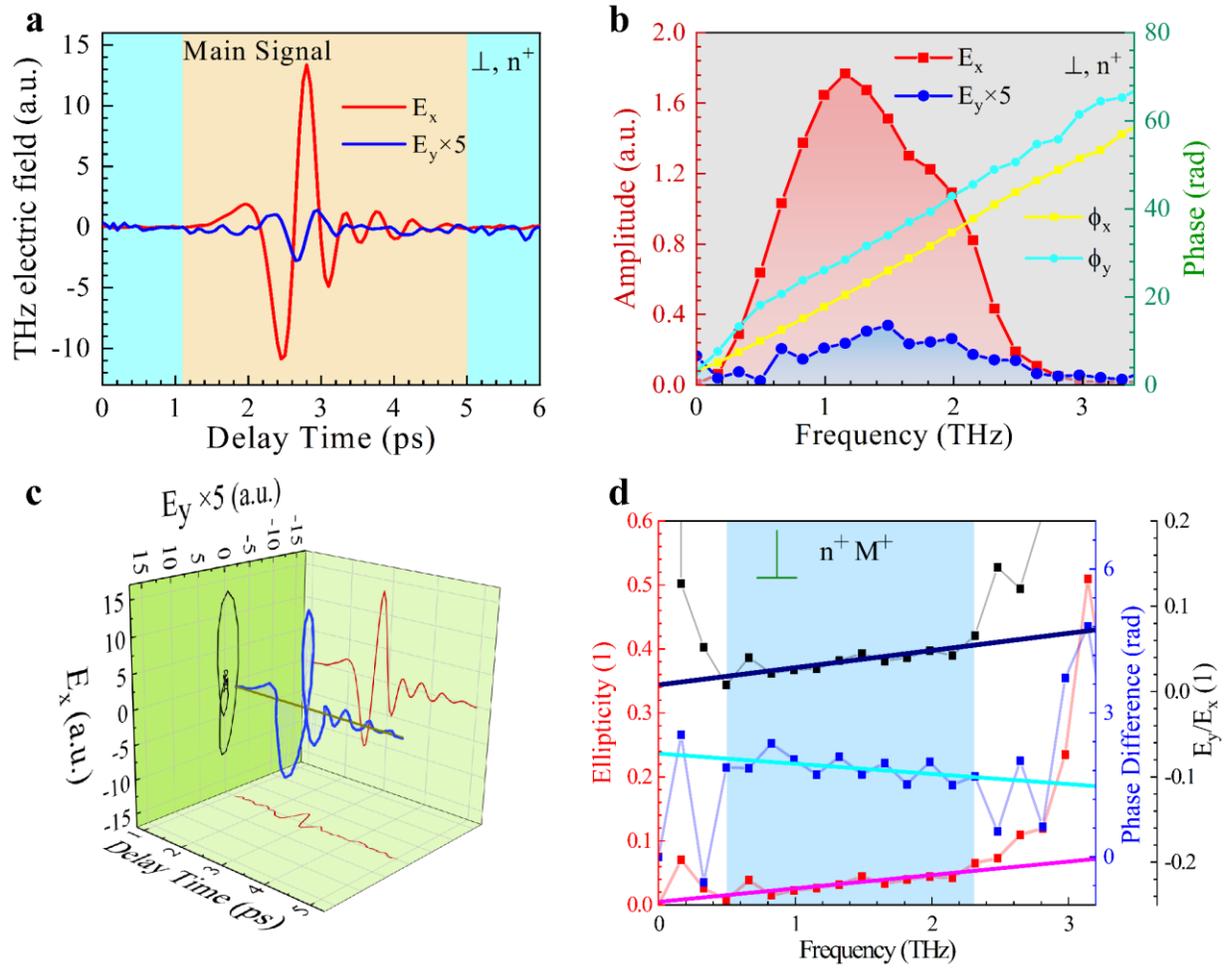

**FIG. 5**

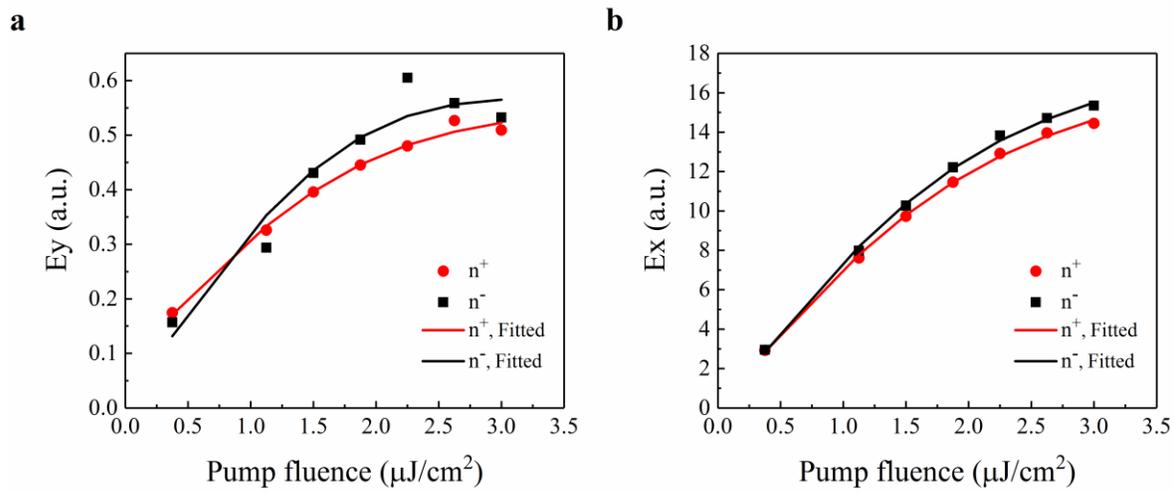

**FIG. 6**